\documentclass{emulateapj}

\shorttitle{Pop. III stars from turbulent fragmentation at $z\sim 11$}
\shortauthors{Prieto et al.}

\begin{document}

\title{Pop. III stars from turbulent fragmentation at redshift  $\sim11$}

\author{Joaquin Prieto\altaffilmark{1,2}, Paolo Padoan\altaffilmark{1}, Raul Jimenez\altaffilmark{1}, Leopoldo Infante\altaffilmark{2}}

\altaffiltext{1}{ICREA \& ICC, University of Barcelona (IEEC-UB), Marti i Franques 1, E08028, Barcelona, Spain}
\altaffiltext{2}{Centro de Astro-Ingenier'a \& Departamento de Astronom\'{i}a y Astrof\'{i}sica, Pontificia Universidad Cat\'{o}lica de Chile, Vicu\~{n}a Mackenna 4860, 7820436 Macul, Santiago, Chile}

\begin{abstract}
We report results from a cosmological simulation with non-equilibrium chemistry of 21 
species, including H$_2$, HD, and LiH molecular cooling. Starting from cosmological 
initial conditions, we focus on the evolution of the central 1.8~Kpc region of a 
$3\times10^7$~M$_{\odot}$ halo. The crossing of a 
few $10^6$~M$_{\odot}$ halos and the gas accretion through larger scale filaments 
generate a turbulent environment within this region. Due to the short cooling time caused by the 
non-equilibrium formation of H$_2$, the supersonic turbulence results in a very 
fragmented mass distribution, where dense, gravitationally unstable clumps emerge from 
a complex network of dense filaments. At $z=10.87$, we find approximately 25 well defined, 
gravitationally unstable clumps, with masses of $4\times10^3-9\times10^5$~M$_{\odot}$, 
temperatures of approximately 300~K, and cooling times much shorter than the free-fall time. 
Only the initial phase of the collapse of individual clumps is spatially resolved in the simulation.
Depending on the density reached in the collapse, the estimated average Bonnor-Ebert masses
are in the range $200-800$~ M$_{\odot}$. We speculate that each clump may further fragment into 
a cluster of stars with a characteristic mass in the neighborhood of 50~M$_{\odot}$. This process at 
$z\approx11$ may represent the dominant mode of Pop. III star formation, causing 
a rapid chemical enrichment of the protogalactic environment. 
\end{abstract}


\section{Introduction}

The primordial (Pop. III) star formation process has been extensively studied in the last decade by a 
number of authors \citep{Abeletal2000,Abeletal2002,Brommetal1999,Brommetal2002,Yoshidaetal2006,McGreerBryan2008},
who have concluded that primordial stars of masses in the range $\sim10^2-10^3$M$_{\odot}$ 
are formed in dark matter (DM) mini halos of $\sim10^6$M$_{\odot}$ at a redshift of $z\sim 30-15$. Although in the collapse 
of such mini halos turbulent motions seem not to be important, at least on scales comparable to their virial radius, they are 
important for larger halos and may play a key role in the star formation process at high redshift, as in present-day star formation \citep{Padoanetal2007}. 

\citet{WiseAbel2007} and \citet{Greifetal2008} have discussed the generation of turbulent motions in primordial gas 
through the virialization process of haloes of mass $\sim 10^7$M$_{\odot}$. They have shown that the collapse of 
primordial gas on such halos, and particularly the cold accretion through filaments, generate supersonic turbulent motions,
which partially ionize the primordial gas allowing an efficient formation of H$_2$ and HD molecules and efficient cooling
to a gas temperature of $\sim200-300$K. They have argued that these turbulent, low temperature regions could be sites of 
efficient star formation. However, their simulations do not spatially resolve the fragmentation process to the scale of individual
dense clumps associated with star-forming regions, and adopt adaptive mesh refinement (AMR) and smooth particle 
hydrodynamics (SPH) methods that were not specifically designed to resolve well the turbulent cascade within the central 
1.8~Kpc region. 

In this letter, we present the results of an N-body and hydrodynamical simulation with non-equilibrium chemistry and 
cooling (including H$_2$, HD, and LiH cooling), carried out with the cluster Geryon at the AIUC-PUC.
Starting from cosmological initial conditions, the simulation focuses 
on the evolution of the central 1.8~Kpc region of a $3\times10^7$~M$_{\odot}$ halo. A uniform mesh of 512$^3$ computational 
cells in that region provides a uniform spatial resolution of 3.5~pc, which allows us to follow the generation of turbulent 
motions in the primordial gas by the combined effects of the crossing of a few $10^6$~M$_{\odot}$ halos and of the gas 
accretion through larger scale filaments. The turbulence results in the formation of a large number of gravitationally 
unstable clumps at $z\approx11$, with a characteristic mass of $10^5$~M$_{\odot}$. These clumps may be sites of 
efficient star formation in primordial gas, well before the virialization process of the first galaxies.

\begin{figure*}[t]
\includegraphics[scale=.34]{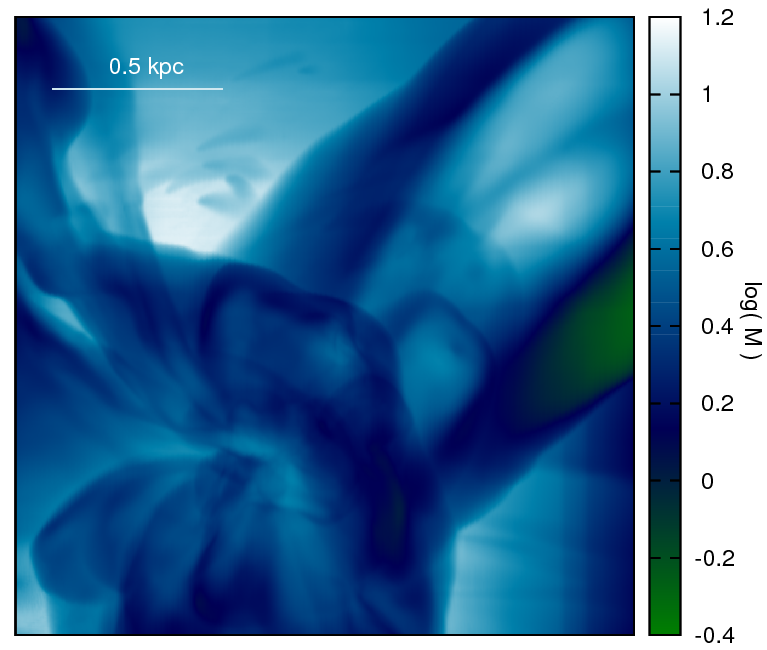}
\includegraphics[scale=.34]{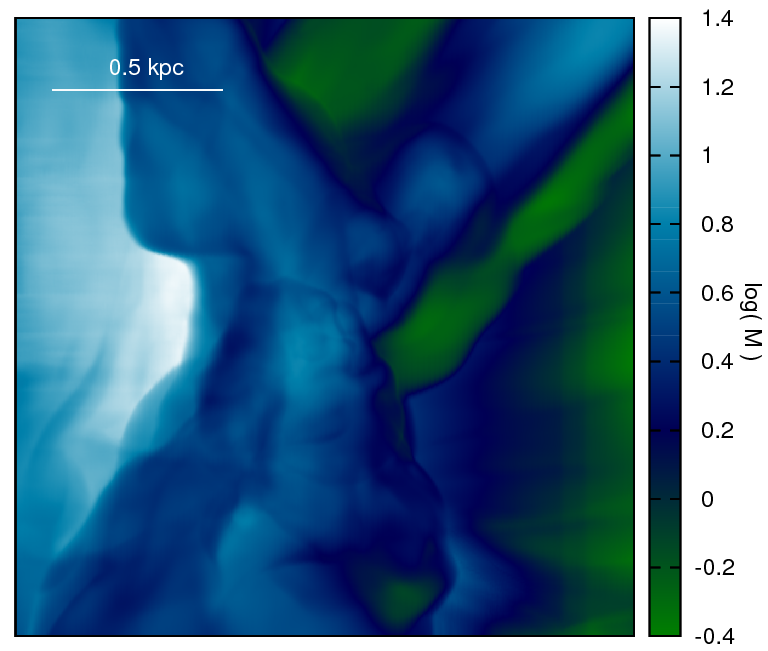}
\includegraphics[scale=.34]{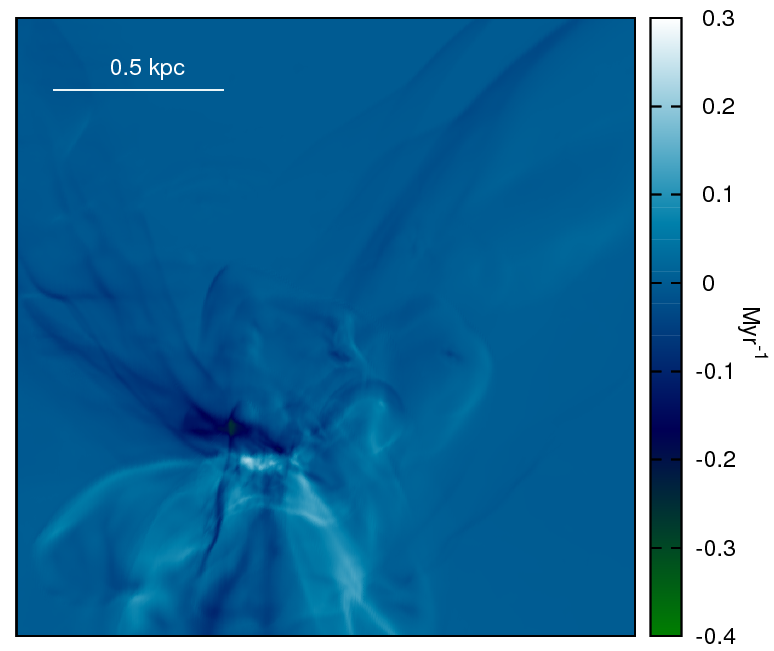}
\includegraphics[scale=.34]{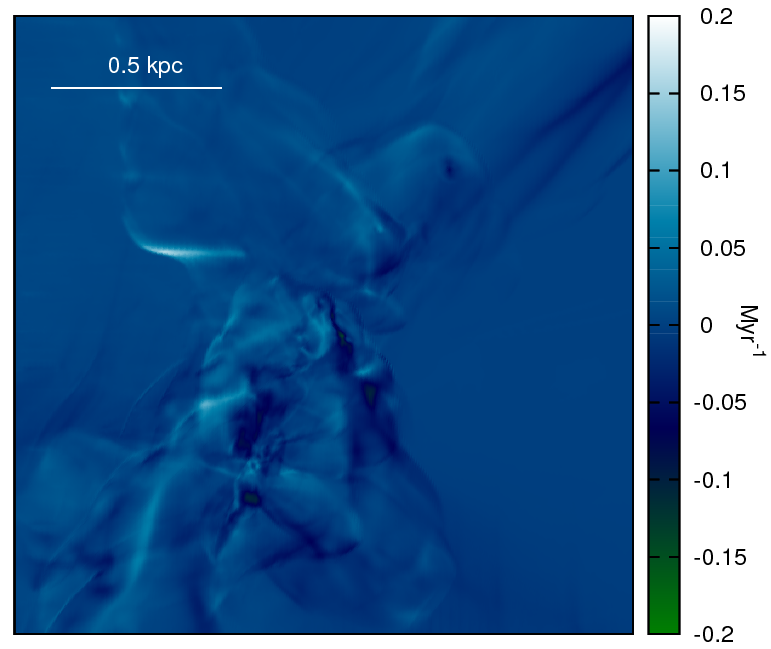}
\caption{Upper panels: Line-of-sight gas rms velocity, averaged along the $y$ direction (left) and $z$ direction (right), 
divided by the sound speed averaged along the line of sight. Only the central region of the simulation is shown, 
where the uniform resolution mesh of 512$^3$ zones covers a size of 1.8~Kpc at $z=10.87$. Lower panels: 
Average value of the voriticity component parallel to the line of sight for the same regions shown in the top panels.}
\label{f1}
\end{figure*}

\section{Methodology}

We used the AMR code RAMSES
\citep{Teyssier2002} with a modified non-equilibrium cooling module with 21 species 
(including H$_2$, HD, and LiH molecules with their cooling functions) in order to follow
the chemo-thermal evolution and the gravitational collapse of primordial gas. All chemical 
reaction rates were taken from \citet{Stanciletal1996}, \citet{GP98}, and \citet{GA08}.

We worked in a concordance $\Lambda$CDM cosmological model: $h=0.72$,
$\Omega_{\Lambda}=0.73$,
$\Omega_{m}=0.27$, $\Omega_{b}=0.04$, $\sigma_8=0.9$ and $n_s=0.95$. The
dynamical initial conditions
were taken from mpgrafic \citep{Prunetetal2008} and the initial chemical
abundances were taken from \citet{GP98}
at $z=120$ (the initial redshift of the simulation).

We first simulated a volume of 1 (Mpc/h)$^3$ of only DM, with 256$^3$ particles. Using the HOP algorithm
\citep{EisensteinHut1998}, we identified a $3\times 10^7$~M$_{\odot}$ halo at $z=10$. In order to better resolve the 
formation of that halo and to study its baryon component, we re-simulated the same 1 (Mpc/h)$^3$ volume with 
512$^3$ particles (with a particle mass of approximately 800~M$_{\odot}$), including gas with non-equilibrium primordial 
chemistry from the beginning of the simulation, at z=120. The gas dynamics was computed on a root grid of 64$^3$
computational zones, which was geometrically refined towards the center, increasing the spatial resolution by a factor of
two within the central 1/8 of the volume. This refinement criterion was applied three times, generating 4 nested meshes
(including the root grid), each with 64$^3$ computational cells. 

Inside the innermost 64$^3$ mesh, the gas dynamics was computed using 6 extra levels of adaptive mesh refinement,
according to four different criteria: i) Lagrangian refinement based on the number density of DM particles (a finer refinement level 
is created in cells containing more than 4 DM particles), ii) Lagrangian refinement based on the baryonic mass density,  iii) refinement 
based on the gas pressure gradient (for $\Delta p/p\ge2$), and iv) refinement based on the Jeans' length, to satisfy 
Truelove's condition \citep{Trueloveetal1997}. The pressure gradient criterion was included in order to better resolve 
the turbulent flow, as discussed in \citep{Kritsuketal2006}. The three geometrical refinement levels, plus the additional 
6 AMR levels, give an effective spatial resolution corresponding to that achieved by a uniform mesh 
of $32,768^3$ computational elements, and corresponding to a proper size of 3.5~pc at $z=10.87$.

This setup was used from the beginning of the simulation, at $z=120$, until $z=50$. At $z=50$, another geometrical refinement 
criterion was added (while maintaining the others): Uniform spatial resolution corresponding to the highest refinement level 
was imposed in a volume centered approximately around the densest region of the halo, 
creating a uniform mesh of 512$^3$ elements. This uniform mesh,
covering a proper size of 1.8~Kpc at $z=10.87$, served the purpose of better resolving the generation of turbulent motions in the 
central region of the halo. The results presented in this work are based on the analysis of this central 1.8~Kpc region at the
end of the simulation, corresponding to $z=10.87$.

\section{Supersonic Turbulence}

At a redshift of $z=10.87$, gas turbulence is well developed in the central 1.8~Kpc region of the halo, and its supersonic
motions have already created a complex network of dense filaments and clumps. The turbulence is driven by gas accretion 
from various large-scale filaments (extending well outside of the 1.8~Kpc region) and by the crossing of a few smaller DM 
halos, with masses of approximately $10^6$~M$_{\odot}$. The upper panels of Figure~1 show the rms line-of-sight velocity 
divided by the mean sound 
speed for each line of sight.  The Mach number associated with this one-dimensional component of the velocity 
is typically $M_{\rm s}\sim 3$, so the three-dimensional turbulent velocity has a characteristic  value of $M_{\rm s} \sim 5$. 
The supersonic turbulence creates a complex 
network of shocks, where vortical motion is generated. The lower panels of Figure~1 show the average 
value of the vorticity component parallel to the line of sight. One can see small-scale motion with characteristic 
timescales of 10 Myr.

Although the large-scale velocity field is driven primarily by the gravitational potential of the DM and baryons, the 
complexity of the collapse geometry appears to be sufficient to feed a turbulent cascade. 
A Helmholtz decomposition of the velocity field shows 
that the kinetic energy associated with solenoidal motions ($\nabla\cdot {\bf u}=0$) is approximately three times larger than 
the kinetic energy of compressible motions ($\nabla \times {\bf u}=0$), not too different from what is found in the inertial range 
of scales of isothermal turbulence driven by an incompressible force \citep{Padoan+Nordlund03,Federrath+09,Kritsuk+09}. 

\begin{figure}[t]
\includegraphics[scale=.6]{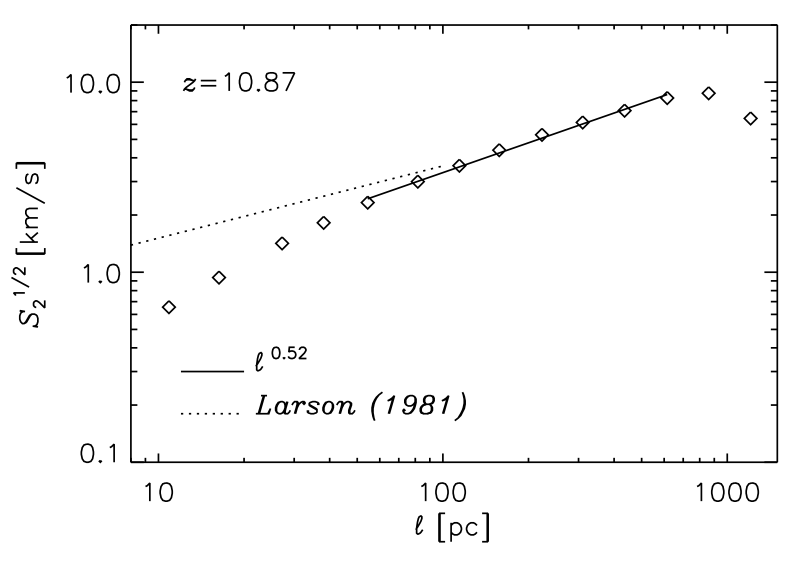}
\caption{Square root  of the second order longitudinal velocity structure function at redshift $z=10.87$. The solid line shows 
the power-law fit in the range of scales (60-600~pc) where the structure function is well approximated by a power law. The dotted
line corresponds to the size-velocity relation found by Larson (1981) in local Galactic molecular clouds.}
\label{f2}
\end{figure}

To further characterize the turbulence, we compute the second order velocity structure function.
The velocity structure function of order $p$ is defined as: 
\begin{equation}
S_p(\ell)=\langle | u({\it \bf x}+{\bf \ell})-u({\it \bf x})|^p\rangle \propto \ell^{\zeta(p)}
\label{sf}
\end{equation}
where the velocity component $u$ is parallel (longitudinal structure function) or perpendicular (transversal structure function) 
to the vector $\bf \ell$, the spatial average is over all values of the position $\bf x$, and $\zeta(p)$ is the exponent of a power law
fit to the structure function. The second order longitudinal structure function averaged in the central region of the simulation
at $z=10.87$ is plotted in Figure~2. It is well approximated by a power law, $\ell^{1.04}$, in the range of scales $\ell =60-600$~pc.
At smaller scales, the structure function becomes steeper, due to numerical diffusion; at larger scales it reaches a peak, probably
indicative of a characteristic outer scale of approximately 600~pc, above which the filamentary gas accretion does not generate 
turbulent motions (consistent with the Mach number and vorticity maps of Figure~1). The slope of the structure function, 
$\zeta(2)=1.04$, is a bit steeper than the value of $\zeta(2)=0.95$ found in high-resolution simulations of randomly driven 
supersonic turbulence \citep{Kritsuk+07}. However, this can be mostly a consequence of the rather diffusive Ramses solver 
used for the simulation (the Local Lax-Friedrich Riemann solver with `minnmod' slope limiter), which is known to generate 
steeper velocity scaling than less diffusive solvers available in Ramses (but generally unstable with supersonic turbulence 
and self-gravity).

In order to relate this result with the observed properties of local Galactic star-forming regions, Figure~2 actually shows the
square root of the second order structure function, next to the velocity scaling law found in Galactic molecular clouds
\citep{Larson79,Larson81}. Interestingly, on the scale of approximately 100~pc, the velocity dispersion in the primordial gas of
our simulation is comparable to that of the molecular gas in our Galaxy. However, due to the larger temperature in the primordial
gas, the Mach number of the turbulence is a few times smaller than in nearby molecular clouds at the same scale.

\begin{figure*}[t]
\includegraphics[scale=.34]{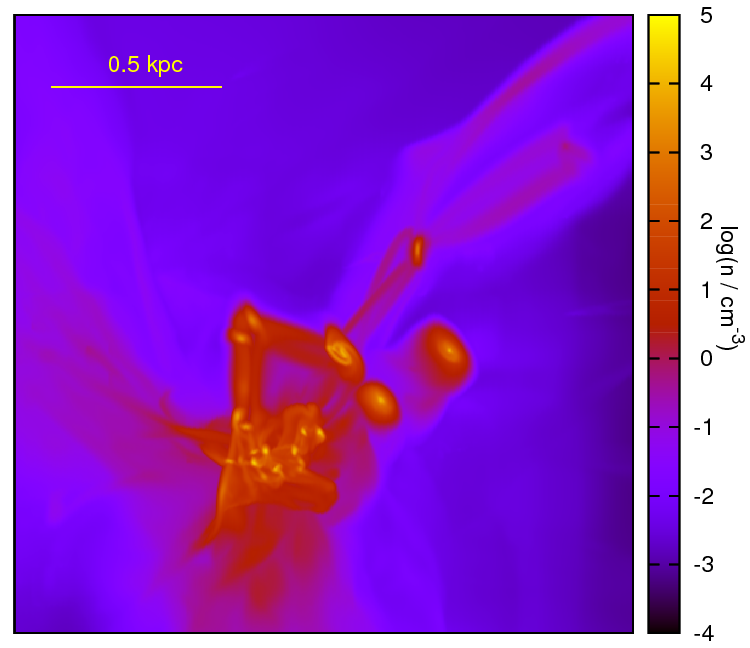}
\includegraphics[scale=.34]{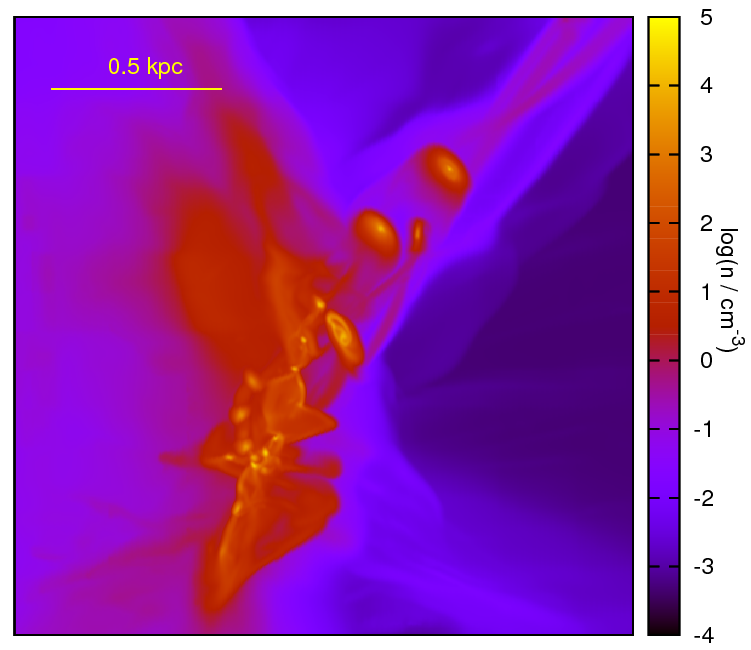}
\includegraphics[scale=.34]{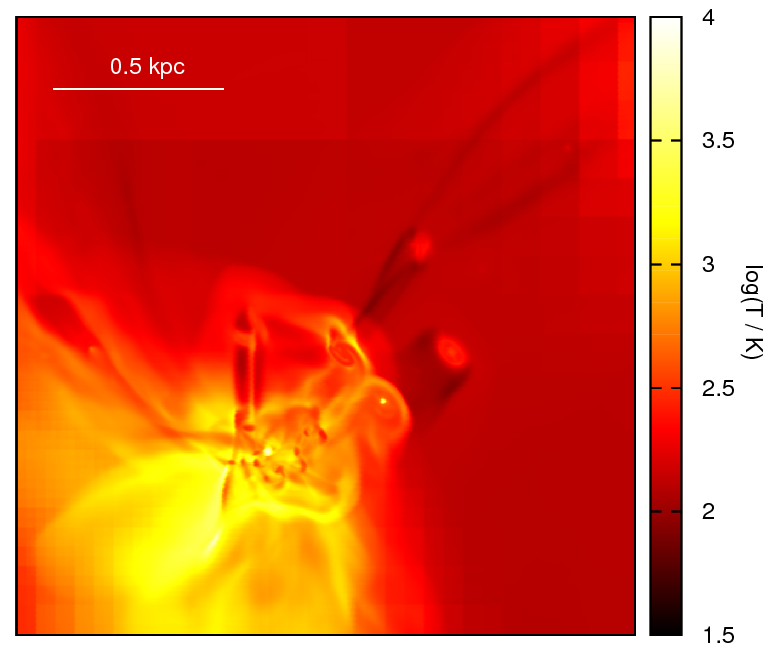}
\includegraphics[scale=.34]{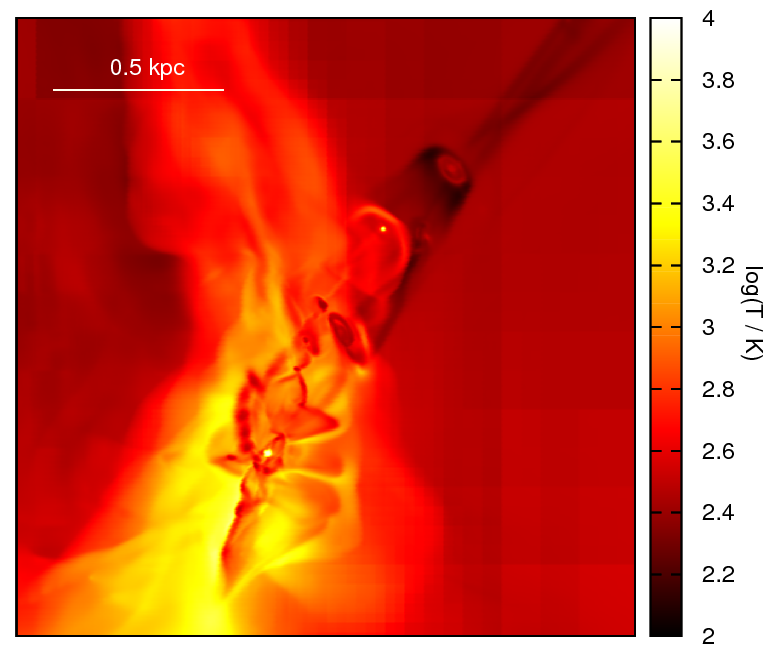}
\caption{Upper panels: Mass weighted gas number density at $z=10.87$, in the central region of the simulation covered by the
uniform resolution mesh of $512^3$ zones, averaged along the $y$ (left panel) and the $z$ (right panel) directions. 
Lower panels: Gas temperature averaged along the line of sight for the same regions shown in the upper panels 
(and in Figure~1). Several well-defined cores can be seen, with $n\sim 10^4$~cm$^{-3}$ and $T\sim 300$~K.}
\label{f3}
\end{figure*}

\section{Gravitationally Unstable Clumps}

Several dense clumps are found at the final redshift of the simulation, $z=10.87$, with only two being clearly related
to local over-densities of DM particles. These clumps are the direct result of the supersonic turbulence generated by
the gas accretion into the central region of the halo. The turbulence creates a network of interacting shocks, forming
dense intersecting filaments. The densest regions of this complex density field may be gravitationally unstable and 
result in collapsing clumps. The upper panels of Figure~3 show the mass weighted gas number density at $z=10.87$, 
averaged along the $y$ (left panel) and the $z$ (right panel) directions. Over 20 well defined clumps can be seen, 
reaching densities above 10$^4$~cm$^{-3}$. As shown in the lower panels of the same Figure~3, these clumps are 
very effectively cooled down to temperatures of approximately 300~K.

In order to study the properties of these clumps, we have selected gravitationally unstable clumps using the 
clumpfind algorithm of \citet{Padoanetal2007}. In this algorithm, clumps are defined as connected overdensities 
that cannot be split into two or more overdensities. The unstable ones are simply those with mass larger than their 
Bonnor-Ebert mass. These definitions are implemented  in the clumpfind algorithm by scanning the density field with
discrete density levels, each a factor $f$ larger than the previous one. Only the connected regions above each 
density level that are larger than their Bonnor-Ebert mass are selected as unstable
clumps. After this selection, the unstable clumps from all levels form a hierarchy tree, of which only the final (unsplit)
clump of each branch is retained.

\begin{table*}[th]
\label{table1}
\caption{Averaged core physical quantities.}
\begin{center}
\begin{tabular}{c c c c c c c c c c c}
\hline\hline
& & & & & & & & &\\
Mass & BE Mass & n & T & n$_{H_2}$/n$_H$ & n$_{HD}$/n$_H$ & n$_{LiH}$/n$_H$ & $\Lambda_{HD}/\Lambda_{H_2}$ & $\Lambda_{LiH}/\Lambda_{H_2}$ & t$_{cool}$ & t$_{cool}$/t$_{ff}$ \\
(M$_{\odot}$) & (M$_{\odot}$) & (cm$^{-3}$) & (K) &  &  &  &  &  & (yr) &  \\
& & & & & & & & & \\
\hline 
& & & & & & & & & \\
8.78 (5)\footnotemark[1] & 3.17 (4) & 3.04 (3) & 6.55 (3) & 6.93 (-4) & 3.77 (-8) & 4.54 (-19) & 4.43 (-4) & 2.38 (-15) & 4.30 (3) & 5.08 (-3) \\ 
7.33 (5) & 3.63 (2) & 2.97 (3) & 3.30 (2) & 1.82 (-2) & 1.87 (-6) & 4.98 (-19) & 4.66 (-2) & 2.66 (-15) & 1.15 (4) & 1.34 (-2) \\ 
7.25 (5) & 2.61 (2) & 3.37 (3) & 2.76 (2) & 2.36 (-2) & 2.19 (-6) & 4.89 (-19) & 6.79 (-2) & 3.94 (-15) & 1.40 (4) & 1.74 (-2) \\
5.81 (5) & 8.17 (3) & 1.54 (3) & 2.11 (3) & 1.65 (-3) & 9.13 (-8) & 2.30 (-19) & 3.29 (-4) & 1.21 (-16) & 2.31 (3) & 1.94 (-3) \\
5.15 (5) & 2.59 (2) & 2.72 (3) & 2.56 (2) & 2.59 (-2) & 3.41 (-6) & 4.08 (-19) & 1.01 (-1) & 3.26 (-15) & 1.62 (4) & 1.82 (-2) \\
3.40 (5) & 4.58 (2) & 1.20 (3) & 2.85 (2) & 7.63 (-3) & 8.53 (-7) & 2.02 (-19) & 4.40 (-2) & 2.30 (-15) & 5.77 (4) & 4.28 (-2) \\
3.02 (5) & 5.84 (2) & 1.53 (3) & 3.63 (2) & 5.70 (-3) & 4.82 (-7) & 3.11 (-19) & 2.16 (-2) & 2.57 (-15) & 3.64 (4) & 3.05 (-2) \\
2.79 (5) & 3.91 (2) & 2.28 (3) & 3.18 (2) & 1.15 (-2) & 1.19 (-6) & 4.02 (-19) & 4.44 (-2) & 3.22 (-15) & 2.21 (4) & 2.26 (-2) \\
2.65 (5) & 3.58 (2) & 2.62 (3) & 3.13 (2) & 1.50 (-2) & 1.48 (-6) & 4.25 (-19) & 4.73 (-2) & 2.99 (-15) & 1.68 (4) & 1.84 (-2) \\
2.49 (5) & 5.12 (2) & 1.08 (3) & 2.96 (2) & 7.27 (-3) & 6.19 (-7) & 1.54 (-19) & 2.91 (-2) & 1.53 (-15) & 5.76 (4) & 4.05 (-2) \\
1.31 (5) & 5.05 (2) & 1.38 (3) & 3.18 (2) & 5.97 (-3) & 5.67 (-7) & 2.34 (-19) & 3.12 (-2) & 2.60 (-15) & 5.16 (4) & 4.11 (-2) \\
1.23 (5) & 4.79 (2) & 1.09 (3) & 2.84 (2) & 4.97 (-3) & 5.85 (-7) & 1.77 (-19) & 4.45 (-2) & 2.96 (-15) & 9.33 (4) & 6.59 (-2) \\
1.13 (5) & 6.06 (2) & 9.42 (2) & 3.17 (2) & 3.34 (-3) & 2.99 (-7) & 1.50 (-19) & 2.49 (-2) & 2.48 (-15) & 1.13 (5) & 7.40 (-2) \\
8.55 (4) & 4.99 (2) & 1.50 (3) & 3.25 (2) & 5.40 (-3) & 5.25 (-7) & 2.70 (-19) & 3.19 (-2) & 3.29 (-15) & 5.20 (4) & 4.32 (-2) \\
6.57 (4) & 6.24 (2) & 6.73 (2) & 2.89 (2) & 2.31 (-3) & 2.70 (-7) & 1.08 (-19) & 3.50 (-2) & 2.95 (-15) & 2.49 (5) & 1.38 (-1) \\
6.29 (4) & 2.42 (2) & 4.63 (3) & 2.92 (2) & 2.69 (-2) & 2.57 (-6) & 6.92 (-19) & 7.40 (-2) & 5.25 (-15) & 9.75 (3) & 1.42 (-2) \\
5.73 (4) & 7.19 (2) & 7.60 (2) & 3.30 (2) & 2.17 (-3) & 2.32 (-7) & 1.91 (-19) & 2.49 (-2) & 3.89 (-15) & 1.74 (5) & 1.03 (-1) \\
5.57 (4) & 3.94 (2) & 6.83 (2) & 2.14 (2) & 3.60 (-3) & 7.78 (-7) & 1.02 (-19) & 1.28 (-1) & 4.75 (-15) & 3.45 (5) & 1.93 (-1) \\
5.48 (4) & 6.36 (2) & 7.65 (2) & 3.05 (2) & 2.51 (-3) & 2.92 (-7) & 1.58 (-19) & 3.23 (-2) & 3.52 (-15) & 1.84 (5) & 1.09 (-1) \\
3.14 (4) & 2.62 (2) & 3.63 (3) & 2.83 (2) & 1.96 (-2) & 2.11 (-6) & 5.57 (-19) & 7.78 (-2) & 5.28 (-15) & 1.53 (4) & 1.98 (-2) \\
2.85 (4) & 7.44 (2) & 6.79 (2) & 3.25 (2) & 1.62 (-3) & 1.80 (-7) & 1.67 (-19) & 2.55 (-2) & 4.51 (-15) & 2.58 (5) & 1.44 (-1) \\
2.21 (4) & 9.70 (2) & 4.84 (2) & 3.47 (2) & 8.72 (-4) & 8.83 (-8) & 1.44 (-19) & 1.81 (-2) & 5.32 (-15) & 5.07 (5) & 2.39 (-1) \\
2.11 (4) & 3.39 (2) & 1.90 (3) & 2.72 (2) & 1.16 (-2) & 1.24 (-6) & 2.86 (-19) & 5.97 (-2) & 3.32 (-15) & 3.59 (4) & 3.35 (-2) \\
1.42 (4) & 7.86 (2) & 5.00 (2) & 3.05 (2) & 1.37 (-3) & 1.67 (-7) & 1.20 (-19) & 2.92 (-2) & 4.16 (-15) & 4.38 (5) & 2.10 (-1) \\
4.20 (3) & 3.47 (2) & 1.70 (3) & 2.66 (2) & 1.06 (-2) & 1.20 (-6) & 2.56 (-19) & 6.23 (-2) & 3.25 (-15) & 4.32 (4) & 3.82 (-2) \\
& & & & & & & & & \\
\hline
\end{tabular}
\footnotetext[1]{A (B)$\equiv$ A$\times 10^B$}
\end{center}
\end{table*}

The result of the clumpfind algorithm may be sensitive to the value of the parameter $f$ (the density
resolution), but in this case the cores are all so well defined that the same clumps are always selected
for a wide range of values of $f$. Results given here are for $f=2.0$. The other parameter of the
algorithm is the minimum density. We have verified that the results have a weak dependence also on this
parameter, for minimum densities in the range 50-500~cm$^{-3}$, because most clumps are very well
defined and relatively isolated overdensities. Results given here are for a minimum density of 400~cm$^{-3}$.

The clumpfind algorithm selects 25 gravitationally unstable clumps. All the dense clumps seen in Figure~3
are therefore gravitationally unstable. They account for 29\% of the total baryonic mass of 
$1.97\times10^7$~M$_{\odot}$ found in the central 1.8~Kpc region (the DM mass in the same region
is $1.89\times10^7$~M$_{\odot}$). The properties of these clumps are listed in Table~1. They are all cold,
with temperatures of approximately 300~K, except for the first and the fourth of the list. The warmest of these
two clumps is very clearly associated with a DM halo of approximately 10$^6$~M$_{\odot}$.
The Table also shows that the clump masses (excluded the first one that is most clearly associated to a mini-halo)
are in the range  $4.2\times10^3-7.3\times10^5$~M$_{\odot}$, while their Bonnor-Ebert masses span the limited
range $200-800$~M$_{\odot}$ (excluding the two warmest clumps). Therefore, the clumps 
contain typically 100 to 1000 Bonnor-Ebert masses, suggesting the possibility of further fragmentation.

Table~1 also shows that H$_2$ is by far the dominant cooling agent, and that the cooling time is always orders
of magnitude shorter than the free-fall time. The clumps are selected here at a single snapshot in time, and thus each
of them is captured at a different phase of its evolution. However, based on the ratio of cooling to free-fall time, 
all the cores are expected to be able to collapse further, unless virialized by their internal fragmentation.

\section{Star Formation in Primordial Gas}

This simulation describes the process of turbulent fragmentation of primordial gas in the central region of a 
$3\times10^7$~M$_{\odot}$ halo. Gravitationally unstable clumps are barely resolved, and we can only speculate
about their fragmentation into stars. Their characteristic mass and densities are not much larger than those of
giant molecular clouds (GMCs) in the Milky Way. Although their internal turbulence is not resolved in the simulation, 
Figure~2 shows that, based on the second order velocity structure function, the velocity dispersion should be comparable
to that of nearby GMCs. Turbulent GMCs with an rms Mach number of $M_{\rm s}\approx20$ form stars with an initial 
mass function (IMF) that peaks at a mass approximately 20 times smaller than the average Bonnor-Ebert mass. This is 
understood as the result of the process of turbulent fragmentation, where regions much denser than the mean are 
created by the supersonic flow, allowing local values of the Bonnor-Ebert mass much smaller than the mean value 
\citep{Padoan+97,Padoan+02,Padoanetal2007}.

The primordial gas in the dense clumps found in the simulation has a characteristic temperature 30 times larger than 
nearby GMCs, so the rms Mach number could be approximately 5 times smaller, $M_{\rm s}\approx4$. This would 
imply a characteristic stellar mass a few times smaller than the average Bonnor-Ebert mass, or approximately 
50~M$_{\odot}$ for the densest clumps.  Assuming a star formation efficiency of 10\%, a $10^5$~M$_{\odot}$ clump of primordial gas
would possibly fragment into a stellar cluster containing approximately 200 stars with masses in the neighborhood 
of 50~M$_{\odot}$ (and many more stars of smaller mass). Because the characteristic stellar mass is some fraction 
of the mean Bonnor-Ebert mass, one cannot rule out the possibility of even smaller values that could arise if the clumps 
could collapse to a much higher density before fragmenting into stars, as illustrated for example in the recent work by \cite{Clark+11}.

The details of this scenario of star formation in primordial gas are very uncertain at this stage, and can only be investigated
with simulations of much larger spatial resolution. However, the results of this simulation shows that approximately one third of the 
primordial gas in the center of halos of a few $10^7$~M$_{\odot}$ is readily converted into gravitationally unstable clumps,
very likely sites of efficient star formation. This process occurring at a redshift of $z\approx11$ may represent the dominant 
mode of Pop.~III star formation, causing a rapid chemical enrichment of the protogalactic environment. Due to this 
early metal enrichment, star formation triggered by the assembly process of larger halos, at $z\approx 2-4$, will 
then be characterized by even shorter cooling times, lower temperatures, and larger turbulence 
Mach numbers, yielding stellar masses already comparable to those of present-day star 
formation environments. 

\acknowledgments

The simulations were performed at the Geryon cluster at PUC. JP thanks Conicyt, Fondap and Mecesup for financial support. RJ and PP than the EU and MICINN for their continuos financial support.



\end{document}